\documentclass[prd,aps,showpacs,superscriptaddress,nofootinbib,eqsecnum,amsfonts,amsmath, 
preprintnumbers,twocolumn,floatfix,floats]{revtex4}

\usepackage{epsfig}
\usepackage{bm}
\usepackage{graphics}
\usepackage{graphicx}
\usepackage{bm}
\usepackage{amssymb}
\usepackage[usenames]{color}

\def\bea{\begin{eqnarray}}
\def\eea{\end{eqnarray}}
\def\vt{\vartheta}

\begin{document}

\newcommand{\rhat}{\hat{r}}
\newcommand{\iotahat}{\hat{\iota}}
\newcommand{\phihat}{\hat{\phi}}
\newcommand{\h}{\mathfrak{h}}
\newcommand{\be}{\begin{equation}}
\newcommand{\ee}{\end{equation}}
\newcommand{\ber}{\begin{eqnarray}}
\newcommand{\eer}{\end{eqnarray}}
\newcommand{\fmerg}{f_{\rm merg}}
\newcommand{\fcut}{f_{\rm cut}}
\newcommand{\fring}{f_{\rm ring}}
\newcommand{\cA}{\mathcal{A}}
\newcommand{\ie}{i.e.}
\newcommand{\df}{{\mathrm{d}f}}
\newcommand{\rmi}{\mathrm{i}}
\newcommand{\rmd}{\mathrm{d}}
\newcommand{\rme}{\mathrm{e}}
\newcommand{\dt}{{\mathrm{d}t}}
\newcommand{\pj}{\partial_j}
\newcommand{\pk}{\partial_k}
\newcommand{\psifl}{\Psi(f; {\bm \lambda})}
\newcommand{\hp}{h_+(t)}
\newcommand{\hc}{h_\times(t)}
\newcommand{\Fp}{F_+}
\newcommand{\Fc}{F_\times}
\newcommand{\Ylm}{Y_{\ell m}^{-2}}
\def\no{\nonumber \\ & \quad}
\def\noQ{\nonumber \\}
\newcommand{\mc}{M_c}
\newcommand{\vek}[1]{\boldsymbol{#1}}
\newcommand{\vdag}{(v)^\dagger}
\newcommand{\btheta}{{\bm \theta}}
\newcommand{\pa}{\partial_a}
\newcommand{\pb}{\partial_b}
\newcommand{\Psieff}{\Psi_{\rm eff}}
\newcommand{\Aeff}{A_{\rm eff}}
\newcommand{\deff}{d_{\rm eff}}
\newcommand{\corr}{\mathcal{C}}
\newcommand{\bvthat}{\hat{\mbox{\boldmath $\vt$}}}
\newcommand{\bvt}{\mbox{\boldmath $\vt$}}

\newcommand{\comment}[1]{{\textsf{#1}}}
\newcommand{\ajith}[1]{\textcolor{magenta}{\textit{Ajith: #1}}}
\newcommand{\sukanta}[1]{\textcolor{blue}{\textit{Sukanta: #1}}}

\newcommand{\AEIHann}{Max-Planck-Institut f\"ur Gravitationsphysik 
(Albert-Einstein-Institut) and Leibniz Universit\"at Hannover, 
Callinstr.~38, 30167~Hannover, Germany}
\newcommand{\WSU}{Department of Physics \& Astronomy, Washington State University,
1245 Webster, Pullman, WA 99164-2814, U.S.A.}
\newcommand{\LIGOCaltech}{LIGO Laboratory, California Institute of Technology, 
Pasadena, CA 91125, U.S.A.}
\newcommand{\TAPIR}{Theoretical Astrophysics, California Institute of Technology, 
Pasadena, CA 91125, U.S.A.}

\title{Systematic errors in measuring parameters of non-spinning compact binary coalescences with post-Newtonian templates}
\preprint{LIGO-P0900296}

\author{Sukanta Bose}
\email{sukanta@wsu.edu}
\affiliation{\WSU}

\author{Shaon Ghosh}
\email{shaonghosh@wsu.edu}
\affiliation{\WSU}

\author{P.~Ajith}
\email{ajith@caltech.edu}
\affiliation{\LIGOCaltech}
\affiliation{\TAPIR}




\begin{abstract}

We study the astrophysical impact of inaccurate and incomplete modeling of the gravitational waveforms from compact binary coalescences (CBCs). We do so by the matched filtering of complete inspiral-merger-ringdown (IMR) signals with a bank of inspiral-phase templates modeled after the 3.5 post-Newtonian TaylorT1 approximant. The rationale for the choice of the templates is three-fold: (1) The inspiral phase of the Phenomenological signals, which are an example of complete IMR signals, is modeled on the same TaylorT1 approximant. (2) In the low-mass limit, where the merger and ringdown phases last much shorter than the inspiral phase, the errors should tend to vanishingly small values and, thus, provide an important check on the numerical aspects of our simulations. (3) Since the binary black hole (BBH) signals are not yet known for mass-ratios above ten and since signals from CBCs involving neutron stars are affected by uncertainties in the knowledge of their equation of state, inspiral templates are still in use in searches for those signals. The results from our numerical simulations are compared with analytical calculations of the systematic errors using the Fisher matrix on the template parameter space. We find that the loss in signal-to-noise ratio (SNR) can be as large as 60\% even for binary black holes with component masses $m_1 = 13M_\odot$ and $m_2 = 20M_\odot$. Also, the estimated total-mass for the same pair can be off by as much as 20\%. Both of these are worse for some higher-mass combinations. Even the estimation of the symmetric mass-ratio $\eta$ suffers a nearly 20\% error for this example, and can be worse than 50\% for the mass ranges studied here. These errors significantly dominate their statistical counterparts (at a nominal SNR of 10). It may, however, be possible to mitigate the loss in SNR by allowing for templates with unphysical values of $\eta$.

\end{abstract}
\maketitle

\section{Introduction}
\label{sec:introduction}

The next (second) generation of gravitational-wave (GW) detectors and their network are being planned to come online around 2014. These detectors include the LIGO detectors \cite{Sigg-LIGOstatus-2008} in Hanford and Livingston, USA, the Virgo detector in Pisa, Italy \cite{VirgoStatus-GWDAW2008}, and the GEO600 detector in Ruthe, Germany \cite{Grote-GEOstatus-2008}. Their quest will not only be to make the first detections but also to begin doing astronomy with them. That entails measuring the source parameters as accurately as possible. Measurement accuracy is critical for triggering searches of electromagnetic counterparts in other observatories, which can provide more complete knowledge about these astrophysical objects by tapping into the complementary channels of information accessible to them. That information in turn will enrich our understanding of how, where, and at what rate these objects are formed, apart from answering other questions and raising new ones. This endeavor will be limited, on the one hand, by the inherent statistical noise in the measurement process and, on the other hand, by the accuracy with which the search templates can model actual gravitational waveforms. The former issue of the statistical measurement error was studied in detail in Ref. \cite{Ajith:2009fz} for estimating the parameters of coalescing binaries of non-spinning black holes. The latter issue is one of systematics and is discussed here for searches using post-Newtonian templates. 

Owing to both the anticipated rates and the theoretical knowledge of their waveforms, coalescing binary black holes (BBHs) are among the most promising GW sources sought by ground-based GW detectors. The inspiral and ringdown phases of these waveforms are accurately known through the post-Newtonian (PN) approximation to General Relativity and black hole perturbation theory, respectively. And developments in numerical relativity (NR) in the last five years has made it possible to compute accurate gravitational waveforms from the hitherto unknown \emph{merger} stage as well~\cite{Pretorius:2005gq,Campanelli:2005dd,Baker05a,Herrmann2006,Sperhake2006,Bruegmann:2006at,Thornburg-etal-2007a,Etienne:2007hr}. In spite of this progress, our knowledge of the gravitational waveform of these and other compact binary coalescences (CBCs) is not complete over regions of the parameter space that can be astrophysically significant. First, numerical merger waveforms from BBH systems are not yet available for mass-ratios greater than ten; also, more studies are awaited on how robust the published merger waveforms for mass-ratios greater than six are. More pertinently, complete inspiral-merger-ringdown (IMR) waveforms are not yet available for mass-ratios above four. Second, for compact binaries involving at least one neutron star (NS) the effects of the variety of possible NS equations of state (EOS) and magnetic field on their waveforms are still under study. Some progress has been made in directly relating the differences among the NS EOS to those in the amplitude and phase evolution of the waveforms \cite{Shibata:2002jb,GondekRosinska:2004eh,Shibata:2007zm,Oechslin:2007gn,Baiotti:2008ra,Liu:2008xy,Yamamoto:2008js,Read:2009yp}. But further research is required to establish the robustness of these results. Regardless of the outcome, unraveling the NS EOS will require a high signal-to-noise ratio (SNR), which may not be very probable in the era of Advanced LIGO or ``AdvLIGO'' (see, e.g., \cite{Read:2009yp} and the references therein). However, that by itself does not imply that the detectability of these systems will be hampered. Indeed, often the hope is expressed that a sufficiently large bank of filters across the parameter space of binary masses (and spins), even if somewhat inaccurate and incomplete, may still be able to capture these waveforms without a significant loss of SNR. Nevertheless, it is critical to establish how big or small that loss is for aiding the formulation of detection strategies for upcoming searches and the planning of future missions.

To quantitatively assess the effects of inaccurate modeling, one needs knowledge of the exact waveforms, which we do not possess. To break this impasse, we employ a strategy that was explored earlier by Cutler and Vallisneri for LISA \cite{Cutler:2007mi}. Namely, we choose a surrogate for the exact waveform that, in spite of its approximate nature, is useful in estimating the loss in SNR and parameter accuracy resulting from our incomplete knowledge of its source.

Specifically, we choose as our surrogate waveforms the Phenomenological inspiral-merger-ringdown waveforms of Ref. \cite{Ajith:2007kx}. 
These are analytic waveforms that have been modeled to have better than 99\% fitting factor with hybrid waveforms constructed from PN and NR waveforms, with mass-ratios from one to four. Here, we take these Phenomenological waveforms to be the ``exact'' waveforms in the extended mass-ratio range from 1 to 8, which is not much wider than their proven range of validity. To assess the drop in SNR suffered from an inaccurate modeling of the CBC waveforms, we compute the fitting-factor of a bank of TaylorT1 3.5PN templates \cite{DIS01} when filtering simulated AdvLIGO data with an IMR signal in it. In the process, we are also able to estimate the error in the maximum likelihood estimates of both mass parameters arising from inaccurate waveform modeling. Similar studies on systematic errors were carried out in Refs. \cite{Farr:2009pg,Buonanno:2009zt}. Those studies differ from ours primarily in that they base their target signals on the effective-one-body (EOB) model \cite{Buonanno:1998gg}. Also, the range of CBC mass pairs for which we numerically compute the loss of SNR and parameter-estimation biases is much wider than those studied in the past. Moreover, we vet our numerical results against analytic Fisher matrix calculations in the parameter region where the two are expected to agree.
As in Ref. \cite{Ajith:2009fz}, here too we limit our attention to the leading harmonic of non-spinning CBC signals. The sensitivity we assume for the second generation detector is that of Advanced LIGO \cite{AdLigoUrl}. 

\section{Binary black hole signals and their parameter estimation}

The gravitational-wave strain in an interferometric detector 
can be expressed in terms of its two linear polarization components
$\hp$ and $\hc$ as 
\be
h(t) = \Fp \hp + \Fc \hc\,,
\ee 
where $\Fp$ and  $\Fc$ are the detector's antenna-pattern functions. 
These functions depend on the two angles locating the source in the 
sky and a third angle specifying the orientation of the polarization
ellipse (see, e.g., Ref. \cite{Pai:2000zt}). 
For transient sources lasting not more than several minutes in a detector's
frequency band, 
such as the ones studied here, these functions can be treated as 
constants in time.

The two polarization components of the dominant harmonic of 
BBH signals studied here are sinusoids with varying amplitude and frequency, 
and have phases $\pi/2$ radians apart relative to each
other. Accordingly, their GW signal in a detector can be written as:
\be
h(t) = C \,  A(t) \,\cos [\varphi(t) + \varphi_0],
\label{eq:hOfTDomMode}
\ee 
where the amplitude coefficient $C$ and the initial phase $\varphi_0$ can be taken 
as constants. The signal's time-varying phase $\varphi(t)$ and amplitude $A(t)$ 
are functions of the physical parameters of the binary, such as the component masses 
(and the spins).

The interferometric data in which these signals are searched for are noisy. Thus, 
any test for establishing the presence of a signal in that data requires the 
modeling of this noise, $n(t)$. We denote its Fourier transform as
\be \label{FTdef}
{\tilde n}(f) = \int_{-\infty}^{\infty} n(t)\, e^{-2\pi \rmi f t} \,\dt \,.
\ee
and take it to be zero-mean Gaussian and stationary:
\begin{mathletters}%
\label{noise}
\bea
{\overline{n(t)}} &=& 0, \label{noisea}\\
{\overline{{\tilde n}^*(f){\tilde n}(f')}} &=& \frac{1}{2}S_{h}(f)\, \delta(f-f')\,. \label{noiseb}
\eea
\end{mathletters}%
Above, the over-bar denotes the ensemble average and $S_{h}(f)$ is the 
Fourier transform of the auto-covariance of the 
detector noise and is termed as its (one-sided) power spectral-density.

We also assume the noise to be additive, which implies that 
when a signal is present in the data $s(t)$, then 
\be
s(t) = h(t) + n(t) \,.
\ee
The noise covariance Eq. (\ref{noiseb}) introduces the following inner-product in the 
function space of signals:
\be\label{innerprod}
\langle a ,\>b \rangle = 4 \Re \int_{0}^{\infty} \! \df\>
{\tilde{a}^* (f) \,\tilde{b}(f) \over S_{h}(f)} \ \ , 
\ee
where $\tilde{a}(f)$ and $\tilde{b}(f)$ are the Fourier
transforms of $a(t)$ and $b(t)$, respectively.
For the above model of the detector noise
the Neyman-Pearson criterion \cite{Helstrom} yields an optimal search statistic,
which when maximized over the amplitude coefficient $C$, is the cross-correlation 
of the data with a normalized template,
\be\label{crosscor}
\rho \equiv \langle \hat{h},s\rangle \ \ ,
\ee
where the normalized template is $\hat{\tilde h}(f) \equiv {\tilde h}(f)/\sqrt{\langle h,\>h \rangle} $. Here,  $\hat{\tilde h}(f)$ obeys the condition 
\be\label{unitnorm}
\| \hat{h}\|^2 \equiv \langle \hat{h},\>\hat{h}\rangle = 1 
\ee
and is said to have a unit-norm. When $s$ is replaced by its unit-norm counterpart (i.e., by $s/\| s\|$) in Eq. (\ref{crosscor}), the resulting inner-product is called the {\em match} between two unit-norm waveforms.

In a ``blind'' search in detector data, where none of the binary's parameters are known \emph{a priori}, 
the search for a GW signal requires maximizing $\rho$ over a ``bank'' of templates (see, for 
e.g.,~\cite{Cokelaer:2007kx}) corresponding to different values of those physical parameters. Apart from the intrinsic source parameters, the waveform also depends on the (unknown) initial phase $\varphi_0$ and the 
time of arrival $t_0$. Maximization over the initial phase $\varphi_0$ is effected by using two orthogonal 
templates for each combination of the physical parameters~\cite{schutz-91}, and the maximization over 
$t_0$ is attained efficiently with the help of the Fast Fourier Transform (FFT) algorithms~\cite{NRecipes}. 

For modeling the signals, we use the analytical Fourier domain IMR waveforms 
proposed in Reference~\cite{Ajith:2007kx}:
\be
{\tilde h}(f) \equiv {A}_{\rm eff}(f) \, e^{\rmi\Psi_{\rm eff}(f)},
\label{eq:phenWave}
\ee
where the effective amplitude and phase are expressed as:
\begin{widetext}
\ber
{A_{\rm eff}}(f) &\equiv& \frac{M^{5/6}}{\deff\,\pi^{2/3}}\sqrt{\frac{5\,\eta}{24}}\,\fmerg^{-7/6}
\left\{ \begin{array}{ll}
\left(f/\fmerg\right)^{-7/6}   & \textrm{if $f < \fmerg$}\\
\left(f/\fmerg\right)^{-2/3}   & \textrm{if $\fmerg \leq f < \fring$}\\
w \, {\cal L}(f,\fring,\sigma) & \textrm{if $\fring \leq f < \fcut$,}\\
\end{array} \right. \nonumber \\
\Psi_{\rm eff}(f) &\equiv& 2 \pi f t_0 + \varphi_0 + \frac{1}{\eta}\,\sum_{k=0}^{7} 
(x_k\,\eta^2 + y_k\,\eta + z_k) \,(\pi M f)^{(k-5)/3}\,.
\label{eq:phenWaveAmpAndPhase}
\eer
\end{widetext}
In the above expressions,
\be
{\cal L}(f,\fring,\sigma) \equiv \left(\frac{1}{2 \pi}\right) 
\frac{\sigma}{(f-\fring)^2+\sigma^2/4}\,
\ee 
is a Lorentzian function that has a  width $\sigma$, and that is centered
around the frequency $\fring$. The normalization constant,
$w \equiv \frac{\pi \sigma}{2} \left(\frac{f_{\rm ring}}
{f_{\rm merg}}\right)^{-2/3}$, is chosen so as to make 
${A}_{\rm eff}(f)$ continuous across the ``transition'' frequency $f_{\rm ring}$. 
The parameter $f_{\rm merg}$ is the frequency at which the power-law changes 
from $f^{-7/6}$ to $f^{-2/3}$. The \emph{effective distance} to the binary is denoted
by $\deff$, which is related to the luminosity distance $d_L$ by $\deff = d_L/C$.
The phenomenological parameters $\fmerg, \fring, \sigma$ and $\fcut$ 
are given in terms of the total mass $M$ and symmetric mass-ratio $\eta$ of the 
binary as
\ber
\pi M \fmerg &=&  a_0 \, \eta^2 + b_0 \, \eta + c_0  \,, \nonumber \\
\pi M \fring &= & a_1 \, \eta^2 + b_1 \, \eta + c_1  \,, \nonumber \\
\pi M \sigma &= & a_2 \, \eta^2 + b_2 \, \eta + c_2  \,, \nonumber \\
\pi M \fcut  &= & a_3 \, \eta^2 + b_3 \, \eta + c_3. 
\label{eq:ampParams}
\eer
The coefficients $a_j, b_j, c_j,~j=0...3$ and $x_k,y_k,z_k,~k=0,2,3,4,6,7$ are 
tabulated in Table I of Ref.~\cite{Ajith:2007xh}. For component masses $m_{1,2}$, the total mass is $M=m_1 + m_2$ and the symmetric mass-ratio is $\eta = m_1m_2/M^2$. For the discussion here, it helps to remember that for a mass-ratio of $m_1/m_2 = 1$, 4, and 8, one has $\eta=0.25$, 0.16, and $\sim 0.1$, respectively.

\subsection{Parameter measurement errors}
\label{sec:Fisher}

Following Cutler and Vallisneri \cite{Cutler:2007mi}, consider the vector space of possible time-series data from a GW detector $\mathbf{s}$, which is just the discrete counterpart of $s(t)$. Let the time-series corresponding to {\em exact} and {\em approximate} GW signals from CBCs, parameterized by $\theta^\mu$, be denoted by $\{\mathbf{h}_{\rm X}(\theta^\mu)\}$ and $\{\mathbf{h}_{\rm A}(\theta^\mu)\}$, respectively. In general, these two sets will lie on two different submanifolds of this vector space. Consider the simple case where the latter denotes just the inspiral part, {\it sans} the merger and ringdown parts, of the former. Clearly, the latter will lie on a submanifold of the space of the former time-series of the complete waveform. If, additionally, the latter waveform were different from the former also owing to a parameter-dependent normalization factor, then the latter can cease to live on a sub-manifold of the former. 

Next, let an exact signal $\mathbf{h}_{\rm X}(\theta^\mu_\mathrm{tr})$ be present in the detector data,
\be
\mathbf{s} = \mathbf{h}_{\rm X}(\theta^\mu_{\rm tr}) + \mathbf{n} \,.
\ee
Then the template belonging to the approximate waveform family that gives the best fit to the above data will, in general, have parameters $\theta^\mu_{\rm bf}$ that are different from $\theta^\mu_\mathrm{tr}$. If we denote that template as $\{\mathbf{h}_{\rm A}(\theta^\mu_{\rm bf})\}$, the best fit parameters are those that minimize the distance between the signal and the template, namely,
\be
\partial_j \|\mathbf{s} - \mathbf{h}_\mathrm{A}(\theta) \|^2\Big|_{\theta_\mathrm{bf}}
= \langle \partial_j \mathbf{h}_\mathrm{A}(\theta_\mathrm{bf}) 
\Big|\mathbf{s} - \mathbf{h}_\mathrm{A}(\theta_\mathrm{bf}) \rangle = 0\,.
\ee
For small differences between the waveforms, the above expression can be Taylor expanded about $\theta_\mathrm{bf}^i = \theta_\mathrm{tr}^i + \Delta\theta^i$ to show that
\newcommand{\hGR}{\mathbf{h}_\mathrm{X}}
\newcommand{\hAP}{\mathbf{h}_\mathrm{A}}
\newcommand{\tbf}{\theta_\mathrm{bf}}
\newcommand{\ttr}{\theta_\mathrm{tr}}
\begin{widetext}
\begin{equation}
\Delta\theta^i
 =  \Big(\Gamma^{-1}(\tbf)\Big)^{ij} \Big\{ \, \big\langle \partial_j \hAP(\tbf), \, \mathbf{n} \big\rangle
+ \big\langle \partial_j \hAP(\tbf), \, \left[\hGR(\ttr) - \hAP(\ttr) \right]\big\rangle\Big\} \, .
\label{paramFisher}
\end{equation}
\end{widetext}
where $\Gamma_{ij}(\tbf) \equiv \langle \partial_i \hAP(\tbf),\>\partial_j \hAP(\tbf)\rangle$. As was noted in Ref. \cite{Cutler:2007mi}, whereas the first term above gives the statistical contribution to the error in the parameter estimation, the second term gives the systematic one. The relevant term is the one that dominates the other for expected signal strengths in a given detector or network. 

A total of nine parameters characterize the non-spinning 
BBH coalescence signals considered here.
They are the total mass $M$, the symmetric mass-ratio $\eta$, the sky-position
angles $(\alpha,\delta)$, the binary's orientation angles $(\psi, \iota)$,
the luminosity distance $d_L$, the initial (or some reference) phase 
$\varphi_0$, and the time of arrival (or some reference time) $t_0$. 
Here we present results for systematic errors in $M$ and $\eta$, and the fractional
loss of SNR, arising from inaccurate waveform modeling. Systematic errors in the complementary set of parameters, especially, the ones obtainable only with multi-site observations, will be presented elsewhere \cite{ABGNetSystErr}.

\subsection{Numerical simulations}

For numerically computing the matched-filter outputs and maximizing them over the template parameters, we took the complete IMR signals as the exact or target signals, with $m_{1,2}\in [13,104]M_\odot$. Thus, the mass-ratio of the target signals ranged from 1 to 8. 
The template bank is chosen to comprise 3.5PN TaylorT1 waveforms \cite{DIS01}, with mass parameters overcovering the mass-range of the target signals. The choice of the template waveforms is governed by the fact that the inspiral phase of the Phenomenological waveforms is modeled after that PN approximant.
The templates are modeled with $M$ and $\eta$ such that $m_{1,2}\in [5,121]M_\odot$, but always with $\eta \leq 0.25$, which is the physical upper-bound. For these studies, we used the method and the code described in Ref. \cite{Ajith:2009fz}. Only one target signal is present in the data at any given time. 

Gravitational-wave searches for CBC signals in a single interferometric detector use unit-norm templates whereby the signal amplitude is measured from the value of the signal-to-noise ratio. The measured values of other parameters, such as a binary's component masses, $t_0$ and $\phi_0$, are those defining the template that yields the maximum match with the injected signal. These are the same values that maximize the match between a unit-norm template and a unit-norm signal. Given a unit-norm signal, the fitting factor (FF) of a template bank is defined as the maximum match it yields for that signal. The FF is a useful construct because $(1-\rm{FF})$ is the fractional loss of SNR of the target signal when filtered with the chosen template bank, and depends only on the mass parameters of the target signals. Specifically, it is independent of the CBC signal parameters that cannot be measured by a single detector, namely, the distance and the polarization, inclination, and sky-position angles. 

\begin{figure*}[tb]
\centering
\includegraphics[width=6.2in]{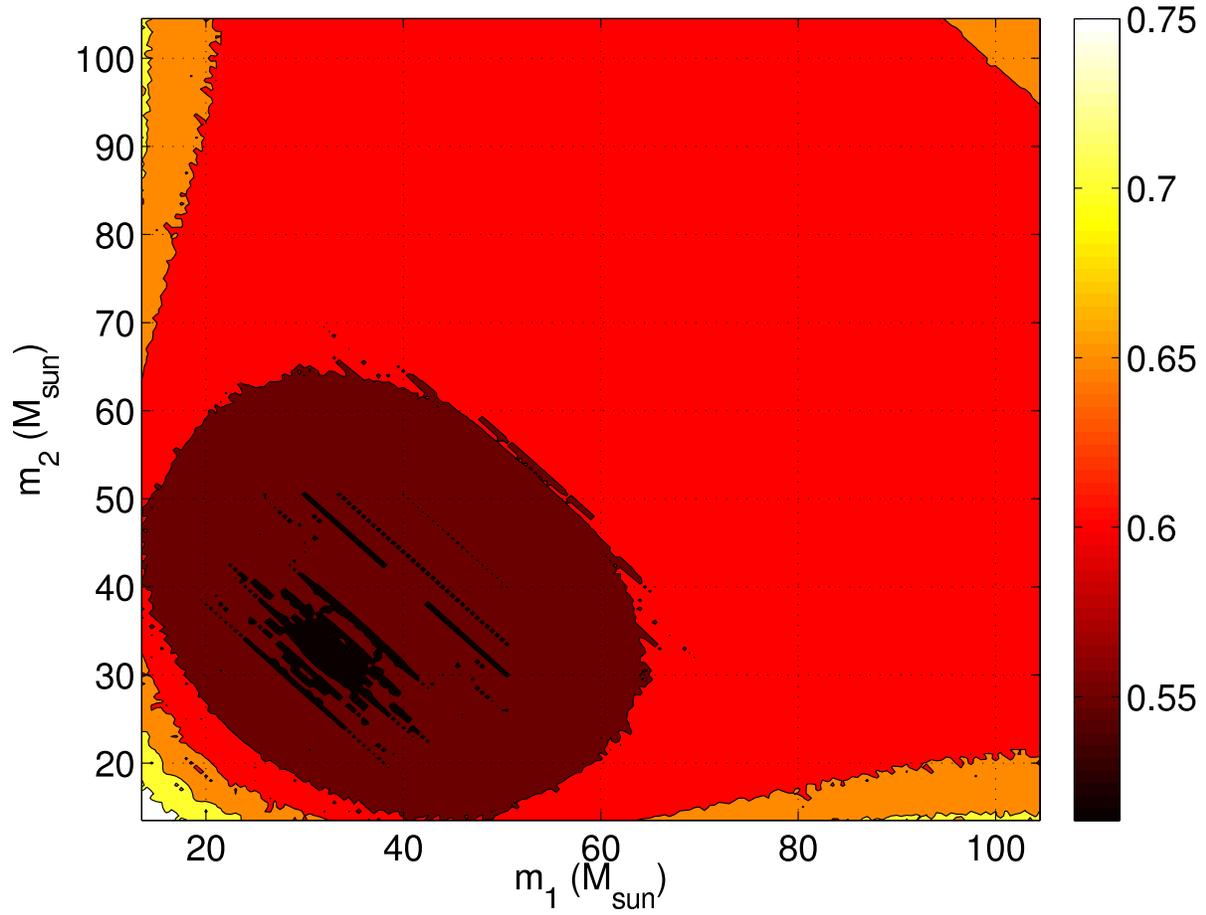}
\caption{The fitting factor obtained from numerical simulations of matched filtering with a bank of 3.5PN TaylorT1 templates of (complete) inspiral-merger-ringdown waveforms as target signals in AdvLIGO PSD. The target waveforms are parameterized by the BBH component masses $m_1$ and $m_2$, each ranging from 13 - 104 $M_\odot$.}
\label{fig:ErrSigmaMonteCarloVsSNR}
\end{figure*}

\begin{figure*}[tb]
\centering
\includegraphics[width=6.2in]{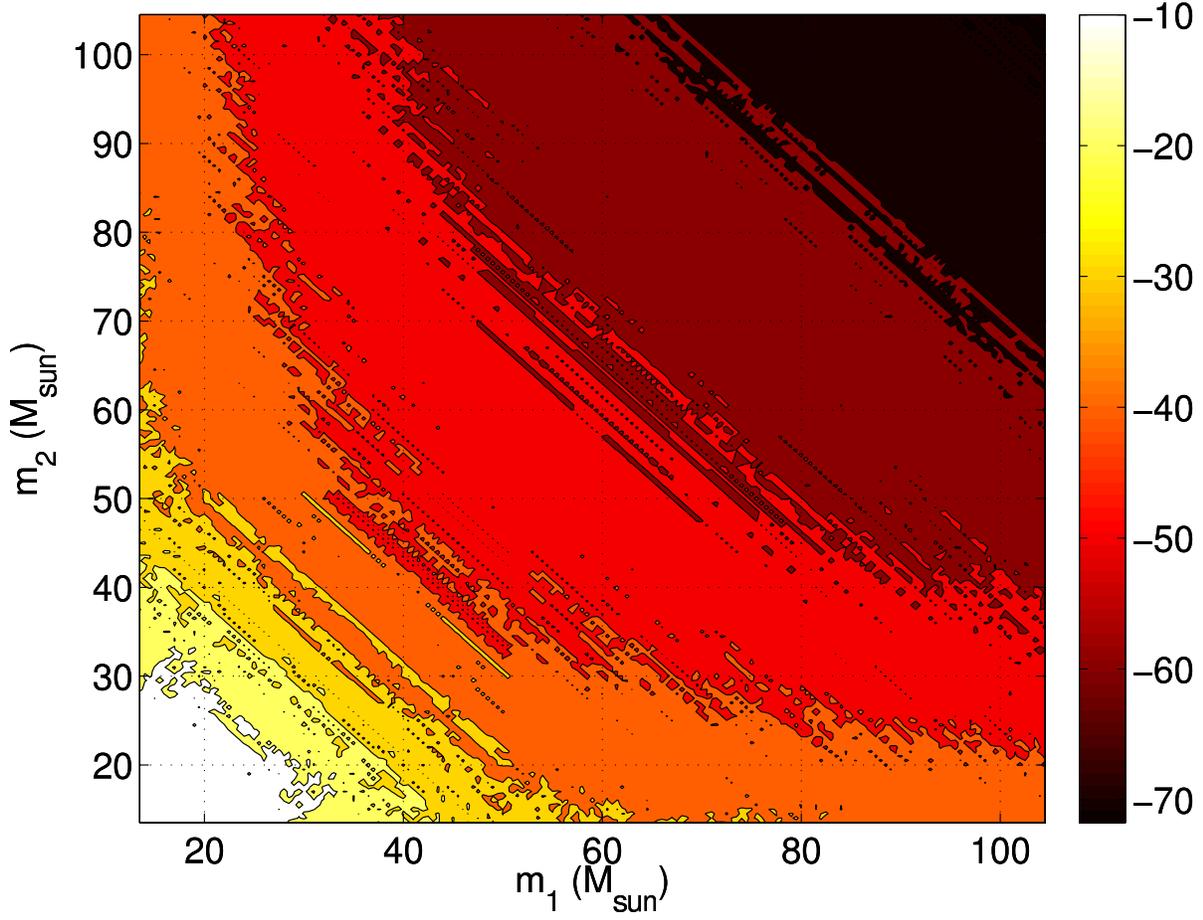}
\caption{The fractional error in total-mass (in \%) obtained from numerical simulations of the same template bank and
target signals, in AdvLIGO PSD, as shown in Fig. \ref{fig:ErrSigmaMonteCarloVsSNR}. Target signals with $\eta=0.25$ are represented by points along the equal-mass line (not shown) extending from the left-bottom corner to the top-right corner of the plot.
}
\label{fig:errTotMassPctMat13To104}
\end{figure*}

\begin{figure*}[tb]
\centering
\includegraphics[width=6.2in]{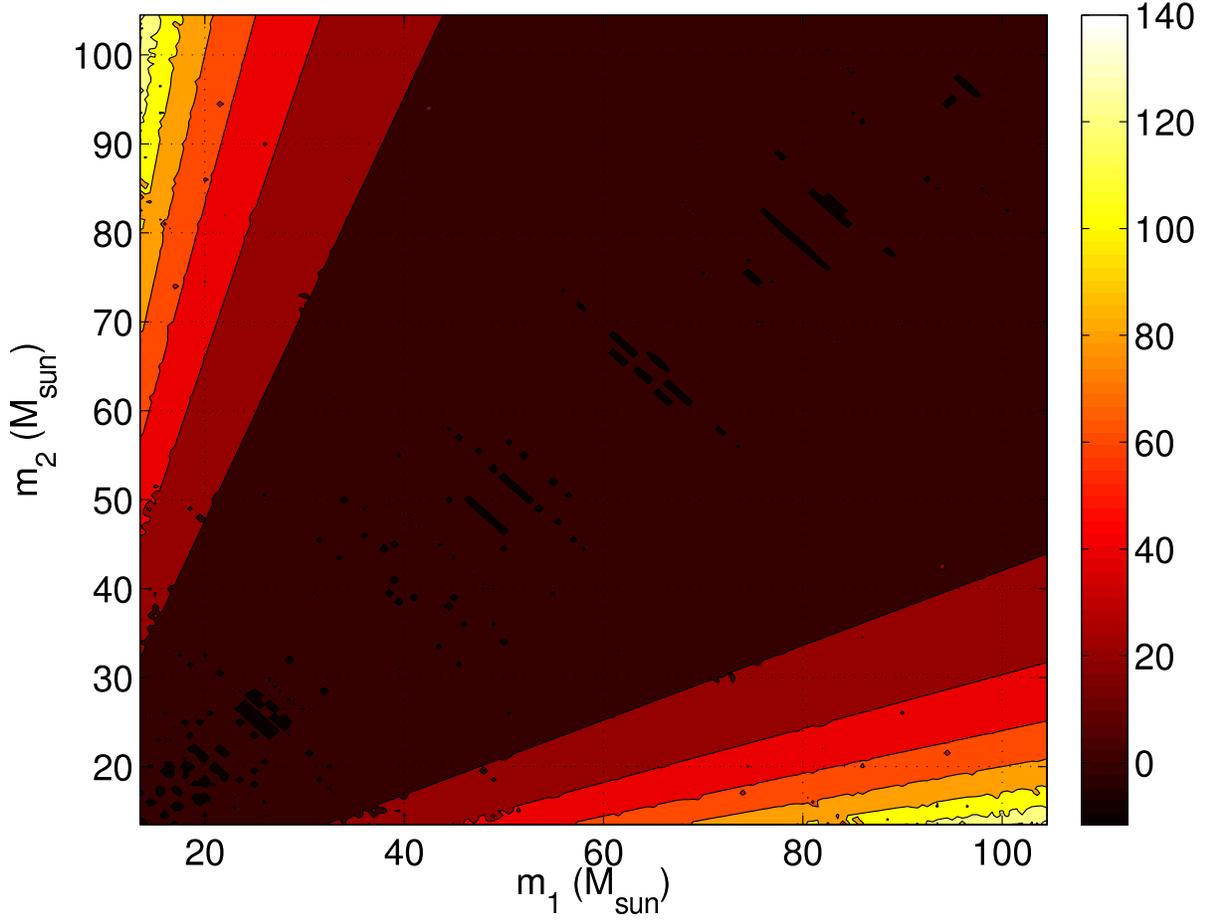}
\caption{The fractional error in the symmetric mass-ratio (in \%)  obtained from numerical simulations for the same template bank and target signals, in AdvLIGO PSD, as shown in Fig. \ref{fig:ErrSigmaMonteCarloVsSNR}.
}
\label{fig:errEtaPctMat13To104}
\end{figure*}

The fitting factor for the above choice of template bank and target signal family is presented in Fig. \ref{fig:ErrSigmaMonteCarloVsSNR}, where the maximization over the template parameters $(t_0,M,\eta)$ was carried out numerically and that over $\phi_0$ was carried out analytically, as explained above. 
Figure \ref{fig:ErrSigmaMonteCarloVsSNR} reveals that for any given mass-ratio, the fitting factor first dips with increasing $M$, before recovering somewhat for higher values of target $M$. This behavior is accompanied by an increasingly negative error in $\Delta M/M$, as shown in Fig. \ref{fig:errTotMassPctMat13To104}, and relatively smaller errors in $\Delta \eta/\eta$, especially, for the region around target $\eta \sim 0.25$, as depicted in Fig. \ref{fig:errEtaPctMat13To104}. This set of observations is explained by the fact that, for any target waveform,
the templates that give the best fit tend to have a smaller $M$, which tends to increase a template's duration, thereby, compensating somewhat its lack of the merger and ringdown phases. As such, for a given template with a fixed $M$, decreasing its $\eta$ increases its duration. So, relative to a target signal's $M$ and $\eta$ values, decreasing a template's $M$ or increasing its $\eta$, or both, can produce waveforms with durations closer to that of the target signal, in principle. In practice, however, the characteristics of the target signals and the modeled templates determine which of these three possibilities is realized.

Note, as well, that for a given target $\eta$, the fitting factor recovers after the dip and improves for large values of target $M$ since it is easier to fit target signals that have a smaller number of cycles. This is consistent with the behavior of the fitting-factor as a function of the total-mass studied in Ref.~\cite{Ajith:2007xh}.
(Similar effects were found in Ref. \cite{Bose:2008ix} when filtering TaylorEt target signals with TaylorT1 and TaylorT4 templates.) We do not present the results for systematic effects on $\phi_0$ since it has little astrophysical relevance and is known to incur large statistical errors (for SNRs around 10). The error in $t_0$ is more interesting, especially, for triggering searches in other electromagnetic observatories, and will be studied in Ref. \cite{ABGNetSystErr}.

\subsection{Analytic approximation using Fisher matrix}

\begin{figure*}[tb]
\includegraphics[height=6.6cm]{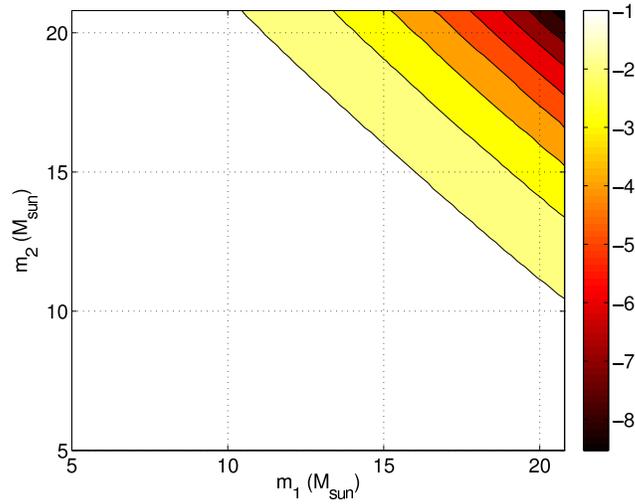}
\caption{The fractional error in total-mass (in \%) given by the analytic expression
Eq. (\ref{paramFisher}) for AdvLIGO PSD. Above, $m_1$ and $m_2$ represent the true parameters.
}
\label{fig:errTotMassPctMat2p5To12p5}
\end{figure*}

\begin{figure*}[tb]
\includegraphics[height=6.6cm]{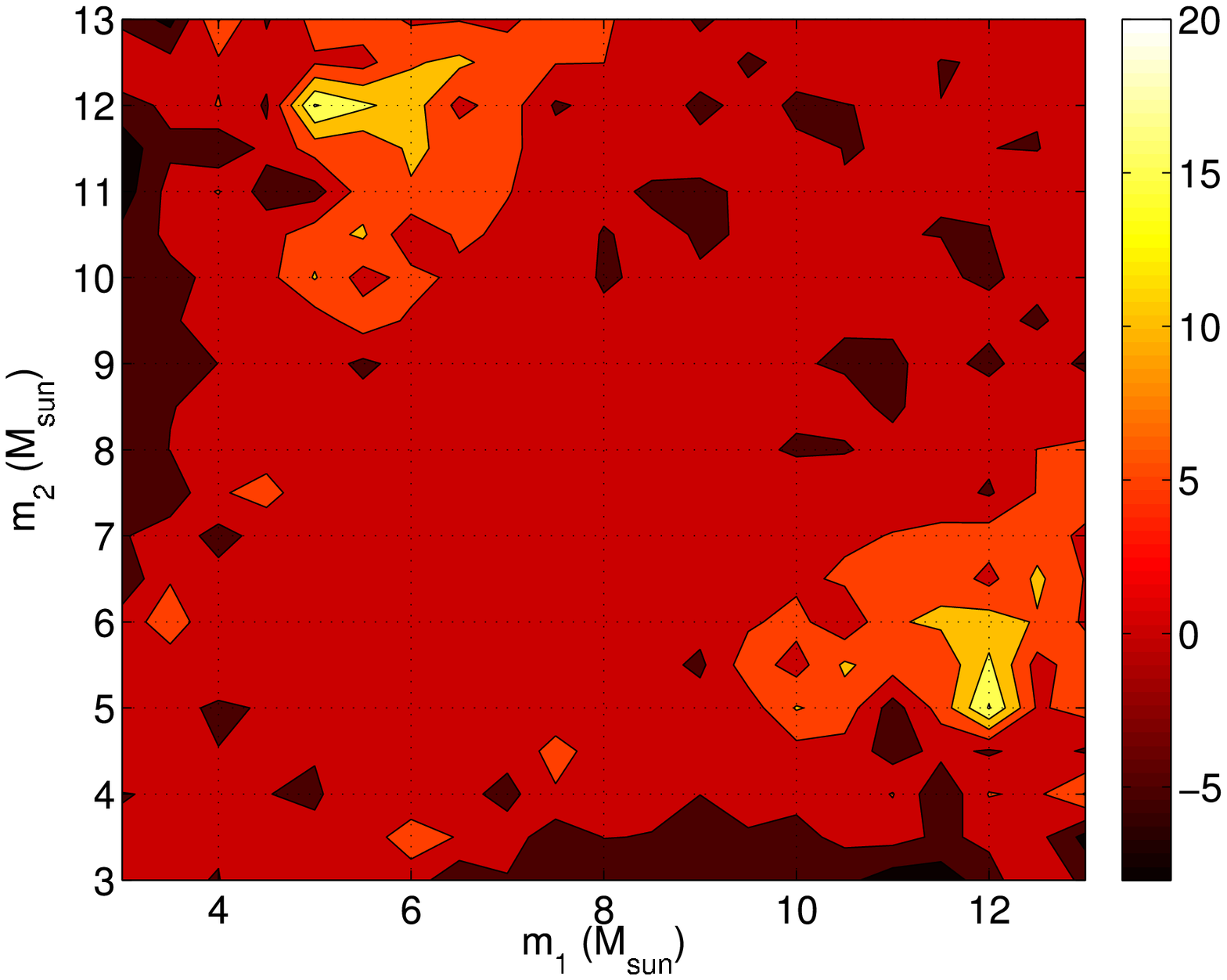}
\includegraphics[height=6.6cm]{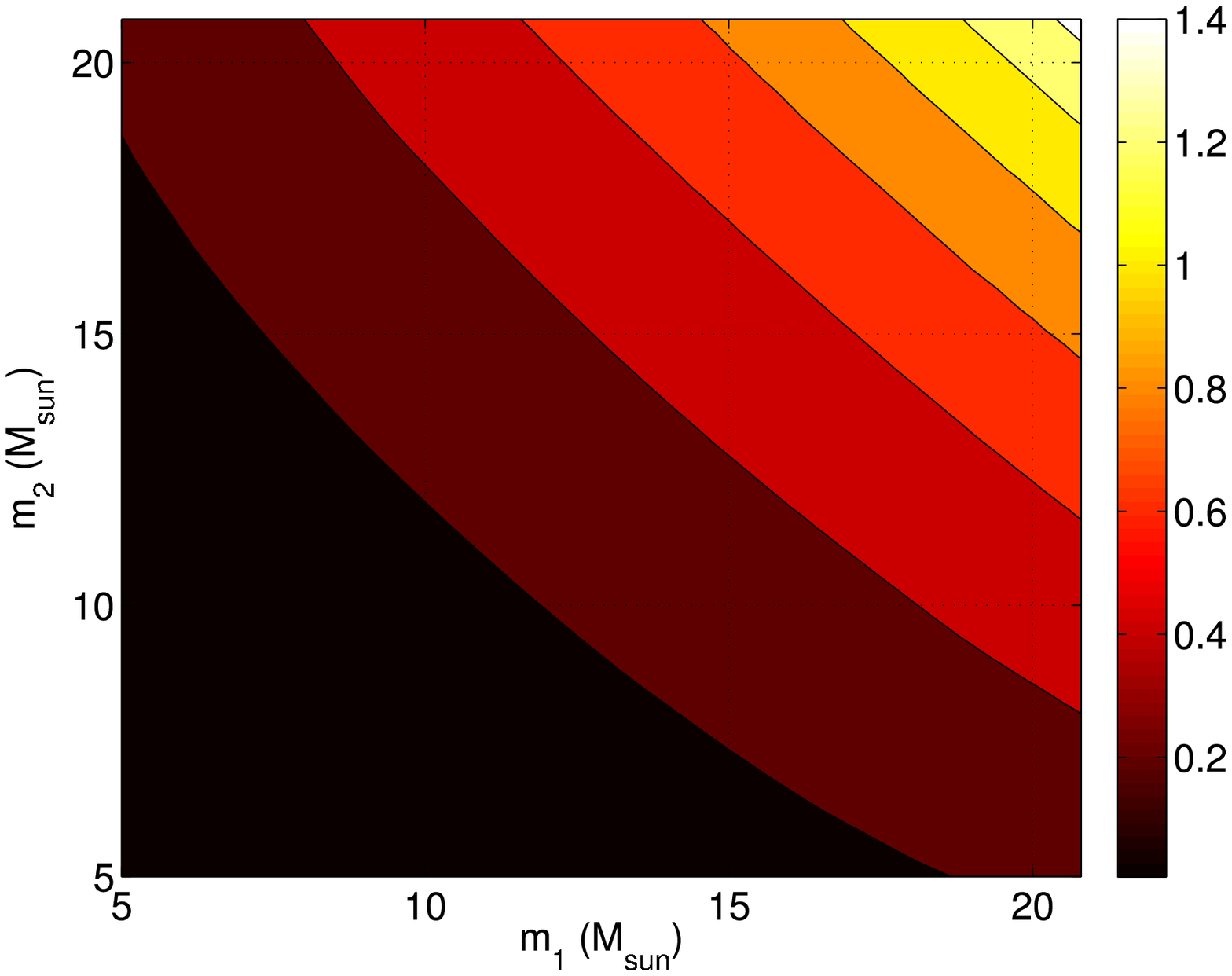}
\caption{The fractional error in the symmetric mass-ratio in AdvLIGO PSD (in \%). The figure on the left is obtained from numerical simulations, and that on the right is computed from the analytic expression in Eq. (\ref{paramFisher}).
}
\label{fig:errEtaPctMat2p5To12p5}
\end{figure*}

We begin by recalling that Eq. (\ref{paramFisher}) was obtained by dropping quadratic and higher order terms in $\Delta\theta^i$ and, thus, is not expected to be valid for large discrepancies in the waveform model. The magnitude of errors found in our numerical studies makes it manifest that the discrepancy between our target signals and templates is smaller for smaller $M$ and higher $\eta$. We, therefore, use the analytic approximation in  Eq. (\ref{paramFisher}) to compute the systematic errors in the parameters in that region, i.e., for $m_{1,2} \in [5-20] M_\odot$. The results are given for $\Delta M/M$ in Fig. \ref{fig:errTotMassPctMat2p5To12p5} and for $\Delta \eta/\eta$ in Fig. \ref{fig:errEtaPctMat2p5To12p5} (in the right plot). Note that we expect the approximation to get worse with increasing target $M$ and decreasing $\eta$. So the primary purpose of these figures is to check if the basic trends seen in the numerical results discussed above are consistent with the theoretical implications of Eq. (\ref{paramFisher}).

Figure \ref{fig:errTotMassPctMat2p5To12p5} nicely corroborates the numerical result presented in Fig. \ref{fig:errTotMassPctMat13To104}, namely, that the longer templates (with smaller $M$) provide a better fit for any target signal and that the magnitude of the fractional error in the estimate of $M$ increases with increasing total mass of the target signal. For $\eta$, the broad trend of templates with larger $\eta$ giving a better fit is visible also in the analytical result. However, the specific feature predicted by Eq. (\ref{paramFisher}) that is at variance with the numerical result is the behavior of  $\Delta \eta/\eta$ near $\eta = 0.25$. Our numerical results show that $\Delta \eta/\eta$ is quite small near $\eta = 0.25$ and does not appear to change even as the target $M$ is increased. Conversely, the analytic formula predicts through the right plot in Fig. \ref{fig:errEtaPctMat2p5To12p5} that the error $\Delta \eta/\eta$ increases with increasing $M$. This apparent disagreement is easily explained by the fact that in our numerical simulations we limited the templates to have $\eta \leq 0.25$; no such restriction is assumed in the derivation of Eq. (\ref{paramFisher}). This is why the best fit value of $\eta$ in the right plot in  Fig. \ref{fig:errEtaPctMat2p5To12p5} is larger than 0.25 as one approaches the top-right corner, i.e., where $m_1 \simeq m_2$ and the target and template waveforms start departing from each other. This also suggests that allowing for templates with unphysical values of $\eta$ might help mitigate some of the loss in SNR arising from inaccurate waveform modeling. (See Ref. \cite{Bose:2008ix} for possible pitfalls of such a strategy.)

\section{Discussion}

In this paper, we studied the loss of SNR and the systematic errors in measuring signal parameters expected in searches for inspiral-merger-ringdown signals from non-spinning BBHs using post-Newtonian templates. Our numerical results show that the fitting factor can be as low as 60\% even for a BBH with $m_1 = 13M_\odot$ and $m_2 = 20M_\odot$. Also, the estimated total-mass for the same pair can be off by as much as 20\%. Both of these estimates get worse for some higher-mass combinations. Even the estimation of $\eta$ suffers a nearly 20\% error for this example, and can be worse than 50\% for the mass ranges studied here. The implications of the loss of SNR and the errors in measuring the masses, such as on inferring the effective distance and other astrophysical properties of the source will be detailed in a future work \cite{ABGNetSystErr}.

\acknowledgments 

Two of us (SB and SG) would like to thank Bruce Allen for his warm hospitality during their stay at Hannover. Some of the computations reported in this paper were performed on the Atlas supercomputing cluster at the Albert Einstein Institute, Hannover. SB also thanks the Kavli Institute for Theoretical Physics, University of California Santa Barbara, and the Aspen Center for Physics, Aspen, Colorado, where early parts of this work were completed. This work is supported in part by NSF grants PHY-0758172, PHY-0855679, PHY-0653653 and PHY-0601459, and the David and Barbara Groce Fund at Caltech. LIGO was constructed by the California Institute of Technology and Massachusetts Institute of Technology with funding from the National Science Foundation and operates under cooperative agreement PHY-0757058.

\bibliography{References}

\begin{thebibliography}{35}
\expandafter\ifx\csname natexlab\endcsname\relax\def\natexlab#1{#1}\fi
\expandafter\ifx\csname bibnamefont\endcsname\relax
  \def\bibnamefont#1{#1}\fi
\expandafter\ifx\csname bibfnamefont\endcsname\relax
  \def\bibfnamefont#1{#1}\fi
\expandafter\ifx\csname citenamefont\endcsname\relax
  \def\citenamefont#1{#1}\fi
\expandafter\ifx\csname url\endcsname\relax
  \def\url#1{\texttt{#1}}\fi
\expandafter\ifx\csname urlprefix\endcsname\relax\def\urlprefix{URL }\fi
\providecommand{\bibinfo}[2]{#2}
\providecommand{\eprint}[2][]{\url{#2}}

\bibitem[{\citenamefont{Sigg}(2008)}]{Sigg-LIGOstatus-2008}
\bibinfo{author}{\bibfnamefont{D.}~\bibnamefont{Sigg}},
  \bibinfo{journal}{Class. Quantum Grav.} \textbf{\bibinfo{volume}{25}},
  \bibinfo{pages}{114041} (\bibinfo{year}{2008}).

\bibitem[{\citenamefont{Acernese et~al.}(2008)}]{VirgoStatus-GWDAW2008}
\bibinfo{author}{\bibfnamefont{F.}~\bibnamefont{Acernese}}
  \bibnamefont{et~al.}, \bibinfo{journal}{Class. Quantum Grav.}
  \textbf{\bibinfo{volume}{25}}, \bibinfo{pages}{184001}
  (\bibinfo{year}{2008}).

\bibitem[{\citenamefont{Grote}(2008)}]{Grote-GEOstatus-2008}
\bibinfo{author}{\bibfnamefont{H.}~\bibnamefont{Grote}},
  \bibinfo{journal}{Class. Quantum Grav.} \textbf{\bibinfo{volume}{25}},
  \bibinfo{pages}{114043} (\bibinfo{year}{2008}).

\bibitem[{\citenamefont{Ajith and Bose}(2009)}]{Ajith:2009fz}
\bibinfo{author}{\bibfnamefont{P.}~\bibnamefont{Ajith}} \bibnamefont{and}
  \bibinfo{author}{\bibfnamefont{S.}~\bibnamefont{Bose}},
  \bibinfo{journal}{Phys. Rev.} \textbf{\bibinfo{volume}{D79}},
  \bibinfo{pages}{084032} (\bibinfo{year}{2009}), \eprint{0901.4936}.

\bibitem[{\citenamefont{Pretorius}(2005)}]{Pretorius:2005gq}
\bibinfo{author}{\bibfnamefont{F.}~\bibnamefont{Pretorius}},
  \bibinfo{journal}{Phys. Rev. Lett.} \textbf{\bibinfo{volume}{95}},
  \bibinfo{pages}{121101} (\bibinfo{year}{2005}), \eprint{gr-qc/0507014}.

\bibitem[{\citenamefont{Campanelli et~al.}(2006)\citenamefont{Campanelli,
  Lousto, Marronetti, and Zlochower}}]{Campanelli:2005dd}
\bibinfo{author}{\bibfnamefont{M.}~\bibnamefont{Campanelli}},
  \bibinfo{author}{\bibfnamefont{C.~O.} \bibnamefont{Lousto}},
  \bibinfo{author}{\bibfnamefont{P.}~\bibnamefont{Marronetti}},
  \bibnamefont{and}
  \bibinfo{author}{\bibfnamefont{Y.}~\bibnamefont{Zlochower}},
  \bibinfo{journal}{Phys. Rev. Lett.} \textbf{\bibinfo{volume}{96}},
  \bibinfo{pages}{111101} (\bibinfo{year}{2006}), \eprint{gr-qc/0511048}.

\bibitem[{\citenamefont{Baker et~al.}(2006)\citenamefont{Baker, Centrella,
  Choi, Koppitz, and van Meter}}]{Baker05a}
\bibinfo{author}{\bibfnamefont{J.~G.} \bibnamefont{Baker}},
  \bibinfo{author}{\bibfnamefont{J.}~\bibnamefont{Centrella}},
  \bibinfo{author}{\bibfnamefont{D.-I.} \bibnamefont{Choi}},
  \bibinfo{author}{\bibfnamefont{M.}~\bibnamefont{Koppitz}}, \bibnamefont{and}
  \bibinfo{author}{\bibfnamefont{J.}~\bibnamefont{van Meter}},
  \bibinfo{journal}{Phys. Rev. Lett.} \textbf{\bibinfo{volume}{96}},
  \bibinfo{pages}{111102} (\bibinfo{year}{2006}), \eprint{gr-qc/0511103}.

\bibitem[{\citenamefont{Herrmann et~al.}(2007)\citenamefont{Herrmann, Hinder,
  Shoemaker, and Laguna}}]{Herrmann2006}
\bibinfo{author}{\bibfnamefont{F.}~\bibnamefont{Herrmann}},
  \bibinfo{author}{\bibfnamefont{I.}~\bibnamefont{Hinder}},
  \bibinfo{author}{\bibfnamefont{D.}~\bibnamefont{Shoemaker}},
  \bibnamefont{and} \bibinfo{author}{\bibfnamefont{P.}~\bibnamefont{Laguna}},
  \bibinfo{journal}{Class. Quantum Gravity} \textbf{\bibinfo{volume}{24}},
  \bibinfo{pages}{S33 } (\bibinfo{year}{2007}).

\bibitem[{\citenamefont{Sperhake}(2006)}]{Sperhake2006}
\bibinfo{author}{\bibfnamefont{U.}~\bibnamefont{Sperhake}}
  (\bibinfo{year}{2006}), \bibinfo{note}{gr-qc/0606079}.

\bibitem[{\citenamefont{Br{\"u}gmann et~al.}(2006)\citenamefont{Br{\"u}gmann,
  Gonz{\'a}lez, Hannam, Husa, Sperhake, and Tichy}}]{Bruegmann:2006at}
\bibinfo{author}{\bibfnamefont{B.}~\bibnamefont{Br{\"u}gmann}},
  \bibinfo{author}{\bibfnamefont{J.~A.} \bibnamefont{Gonz{\'a}lez}},
  \bibinfo{author}{\bibfnamefont{M.}~\bibnamefont{Hannam}},
  \bibinfo{author}{\bibfnamefont{S.}~\bibnamefont{Husa}},
  \bibinfo{author}{\bibfnamefont{U.}~\bibnamefont{Sperhake}}, \bibnamefont{and}
  \bibinfo{author}{\bibfnamefont{W.}~\bibnamefont{Tichy}}
  (\bibinfo{year}{2006}), \bibinfo{note}{gr-qc/0610128}.

\bibitem[{\citenamefont{Thornburg et~al.}(2007)\citenamefont{Thornburg, Diener,
  Pollney, Rezzolla, Schnetter, Seidel, and Takahashi}}]{Thornburg-etal-2007a}
\bibinfo{author}{\bibfnamefont{J.}~\bibnamefont{Thornburg}},
  \bibinfo{author}{\bibfnamefont{P.}~\bibnamefont{Diener}},
  \bibinfo{author}{\bibfnamefont{D.}~\bibnamefont{Pollney}},
  \bibinfo{author}{\bibfnamefont{L.}~\bibnamefont{Rezzolla}},
  \bibinfo{author}{\bibfnamefont{E.}~\bibnamefont{Schnetter}},
  \bibinfo{author}{\bibfnamefont{E.}~\bibnamefont{Seidel}}, \bibnamefont{and}
  \bibinfo{author}{\bibfnamefont{R.}~\bibnamefont{Takahashi}},
  \bibinfo{journal}{Class. Quantum Grav.} \textbf{\bibinfo{volume}{24}},
  \bibinfo{pages}{3911} (\bibinfo{year}{2007}), \eprint{gr-qc/0701038}.

\bibitem[{\citenamefont{Etienne et~al.}(2007)\citenamefont{Etienne, Faber, Liu,
  Shapiro, and Baumgarte}}]{Etienne:2007hr}
\bibinfo{author}{\bibfnamefont{Z.~B.} \bibnamefont{Etienne}},
  \bibinfo{author}{\bibfnamefont{J.~A.} \bibnamefont{Faber}},
  \bibinfo{author}{\bibfnamefont{Y.~T.} \bibnamefont{Liu}},
  \bibinfo{author}{\bibfnamefont{S.~L.} \bibnamefont{Shapiro}},
  \bibnamefont{and} \bibinfo{author}{\bibfnamefont{T.~W.}
  \bibnamefont{Baumgarte}} (\bibinfo{year}{2007}), \eprint{arXiv:0707.2083
  [gr-qc]}.

\bibitem[{\citenamefont{Shibata and Uryu}(2002)}]{Shibata:2002jb}
\bibinfo{author}{\bibfnamefont{M.}~\bibnamefont{Shibata}} \bibnamefont{and}
  \bibinfo{author}{\bibfnamefont{K.}~\bibnamefont{Uryu}},
  \bibinfo{journal}{Prog. Theor. Phys.} \textbf{\bibinfo{volume}{107}},
  \bibinfo{pages}{265} (\bibinfo{year}{2002}), \eprint{gr-qc/0203037}.

\bibitem[{\citenamefont{Gondek-Rosinska et~al.}(2007)}]{GondekRosinska:2004eh}
\bibinfo{author}{\bibfnamefont{D.}~\bibnamefont{Gondek-Rosinska}}
  \bibnamefont{et~al.}, \bibinfo{journal}{Adv. Space Res.}
  \textbf{\bibinfo{volume}{39}}, \bibinfo{pages}{271} (\bibinfo{year}{2007}),
  \eprint{gr-qc/0412010}.

\bibitem[{\citenamefont{Shibata and Taniguchi}(2008)}]{Shibata:2007zm}
\bibinfo{author}{\bibfnamefont{M.}~\bibnamefont{Shibata}} \bibnamefont{and}
  \bibinfo{author}{\bibfnamefont{K.}~\bibnamefont{Taniguchi}},
  \bibinfo{journal}{Phys. Rev.} \textbf{\bibinfo{volume}{D77}},
  \bibinfo{pages}{084015} (\bibinfo{year}{2008}), \eprint{0711.1410}.

\bibitem[{\citenamefont{Oechslin and Janka}(2007)}]{Oechslin:2007gn}
\bibinfo{author}{\bibfnamefont{R.}~\bibnamefont{Oechslin}} \bibnamefont{and}
  \bibinfo{author}{\bibfnamefont{H.~T.} \bibnamefont{Janka}},
  \bibinfo{journal}{Phys. Rev. Lett.} \textbf{\bibinfo{volume}{99}},
  \bibinfo{pages}{121102} (\bibinfo{year}{2007}), \eprint{astro-ph/0702228}.

\bibitem[{\citenamefont{Baiotti et~al.}(2008)\citenamefont{Baiotti, Giacomazzo,
  and Rezzolla}}]{Baiotti:2008ra}
\bibinfo{author}{\bibfnamefont{L.}~\bibnamefont{Baiotti}},
  \bibinfo{author}{\bibfnamefont{B.}~\bibnamefont{Giacomazzo}},
  \bibnamefont{and} \bibinfo{author}{\bibfnamefont{L.}~\bibnamefont{Rezzolla}},
  \bibinfo{journal}{Phys. Rev.} \textbf{\bibinfo{volume}{D78}},
  \bibinfo{pages}{084033} (\bibinfo{year}{2008}), \eprint{0804.0594}.

\bibitem[{\citenamefont{Liu et~al.}(2008)\citenamefont{Liu, Shapiro, Etienne,
  and Taniguchi}}]{Liu:2008xy}
\bibinfo{author}{\bibfnamefont{Y.~T.} \bibnamefont{Liu}},
  \bibinfo{author}{\bibfnamefont{S.~L.} \bibnamefont{Shapiro}},
  \bibinfo{author}{\bibfnamefont{Z.~B.} \bibnamefont{Etienne}},
  \bibnamefont{and}
  \bibinfo{author}{\bibfnamefont{K.}~\bibnamefont{Taniguchi}},
  \bibinfo{journal}{Phys. Rev.} \textbf{\bibinfo{volume}{D78}},
  \bibinfo{pages}{024012} (\bibinfo{year}{2008}), \eprint{0803.4193}.

\bibitem[{\citenamefont{Yamamoto et~al.}(2008)\citenamefont{Yamamoto, Shibata,
  and Taniguchi}}]{Yamamoto:2008js}
\bibinfo{author}{\bibfnamefont{T.}~\bibnamefont{Yamamoto}},
  \bibinfo{author}{\bibfnamefont{M.}~\bibnamefont{Shibata}}, \bibnamefont{and}
  \bibinfo{author}{\bibfnamefont{K.}~\bibnamefont{Taniguchi}},
  \bibinfo{journal}{Phys. Rev.} \textbf{\bibinfo{volume}{D78}},
  \bibinfo{pages}{064054} (\bibinfo{year}{2008}), \eprint{0806.4007}.

\bibitem[{\citenamefont{Read et~al.}(2009)}]{Read:2009yp}
\bibinfo{author}{\bibfnamefont{J.~S.} \bibnamefont{Read}} \bibnamefont{et~al.},
  \bibinfo{journal}{Phys. Rev.} \textbf{\bibinfo{volume}{D79}},
  \bibinfo{pages}{124033} (\bibinfo{year}{2009}), \eprint{0901.3258}.

\bibitem[{\citenamefont{Cutler and Vallisneri}(2007)}]{Cutler:2007mi}
\bibinfo{author}{\bibfnamefont{C.}~\bibnamefont{Cutler}} \bibnamefont{and}
  \bibinfo{author}{\bibfnamefont{M.}~\bibnamefont{Vallisneri}},
  \bibinfo{journal}{Phys. Rev.} \textbf{\bibinfo{volume}{D76}},
  \bibinfo{pages}{104018} (\bibinfo{year}{2007}), \eprint{0707.2982}.

\bibitem[{\citenamefont{Ajith et~al.}(2008)}]{Ajith:2007kx}
\bibinfo{author}{\bibfnamefont{P.}~\bibnamefont{Ajith}} \bibnamefont{et~al.},
  \bibinfo{journal}{Phys. Rev. D} \textbf{\bibinfo{volume}{77}},
  \bibinfo{pages}{104017} (\bibinfo{year}{2008}), \eprint{arXiv:0710.2335
  [gr-qc]}.

\bibitem[{\citenamefont{Damour et~al.}(2001)\citenamefont{Damour, Iyer, and
  Sathyaprakash}}]{DIS01}
\bibinfo{author}{\bibfnamefont{T.}~\bibnamefont{Damour}},
  \bibinfo{author}{\bibfnamefont{B.~R.} \bibnamefont{Iyer}}, \bibnamefont{and}
  \bibinfo{author}{\bibfnamefont{B.~S.} \bibnamefont{Sathyaprakash}},
  \bibinfo{journal}{Phys. Rev. D} \textbf{\bibinfo{volume}{63}},
  \bibinfo{pages}{044023} (\bibinfo{year}{2001}), \bibinfo{note}{erratum-ibid.
  {\bf D}~72 (2005) 029902}, \eprint{gr-qc/0010009}.

\bibitem[{\citenamefont{Farr et~al.}(2009)\citenamefont{Farr, Fairhurst, and
  Sathyaprakash}}]{Farr:2009pg}
\bibinfo{author}{\bibfnamefont{B.}~\bibnamefont{Farr}},
  \bibinfo{author}{\bibfnamefont{S.}~\bibnamefont{Fairhurst}},
  \bibnamefont{and} \bibinfo{author}{\bibfnamefont{B.~S.}
  \bibnamefont{Sathyaprakash}}, \bibinfo{journal}{Class. Quant. Grav.}
  \textbf{\bibinfo{volume}{26}}, \bibinfo{pages}{114009}
  (\bibinfo{year}{2009}), \eprint{0902.0307}.

\bibitem[{\citenamefont{Buonanno et~al.}(2009)\citenamefont{Buonanno, Iyer,
  Ochsner, Pan, and Sathyaprakash}}]{Buonanno:2009zt}
\bibinfo{author}{\bibfnamefont{A.}~\bibnamefont{Buonanno}},
  \bibinfo{author}{\bibfnamefont{B.}~\bibnamefont{Iyer}},
  \bibinfo{author}{\bibfnamefont{E.}~\bibnamefont{Ochsner}},
  \bibinfo{author}{\bibfnamefont{Y.}~\bibnamefont{Pan}}, \bibnamefont{and}
  \bibinfo{author}{\bibfnamefont{B.~S.} \bibnamefont{Sathyaprakash}},
  \bibinfo{journal}{Phys. Rev.} \textbf{\bibinfo{volume}{D80}},
  \bibinfo{pages}{084043} (\bibinfo{year}{2009}), \eprint{0907.0700}.

\bibitem[{\citenamefont{Buonanno and Damour}(1999)}]{Buonanno:1998gg}
\bibinfo{author}{\bibfnamefont{A.}~\bibnamefont{Buonanno}} \bibnamefont{and}
  \bibinfo{author}{\bibfnamefont{T.}~\bibnamefont{Damour}},
  \bibinfo{journal}{Phys. Rev. D} \textbf{\bibinfo{volume}{59}},
  \bibinfo{pages}{084006} (\bibinfo{year}{1999}), \eprint{gr-qc/9811091}.

\bibitem[{AdL()}]{AdLigoUrl}
\emph{\bibinfo{title}{The proposal for advanced ligo is available on-line at}},
  \urlprefix\url{http://www.ligo.caltech.edu/advLIGO/}.

\bibitem[{\citenamefont{Pai et~al.}(2001)\citenamefont{Pai, Dhurandhar, and
  Bose}}]{Pai:2000zt}
\bibinfo{author}{\bibfnamefont{A.}~\bibnamefont{Pai}},
  \bibinfo{author}{\bibfnamefont{S.}~\bibnamefont{Dhurandhar}},
  \bibnamefont{and} \bibinfo{author}{\bibfnamefont{S.}~\bibnamefont{Bose}},
  \bibinfo{journal}{Phys. Rev. D} \textbf{\bibinfo{volume}{64}},
  \bibinfo{pages}{042004} (\bibinfo{year}{2001}), \eprint{gr-qc/0009078}.

\bibitem[{\citenamefont{Helstrom}(1995)}]{Helstrom}
\bibinfo{author}{\bibfnamefont{C.~W.} \bibnamefont{Helstrom}},
  \emph{\bibinfo{title}{Elements of signal detection and estimation}}
  (\bibinfo{publisher}{Prentice-Hall, Inc.}, \bibinfo{address}{Upper Saddle
  River, NJ, USA}, \bibinfo{year}{1995}), ISBN \bibinfo{isbn}{0-13-808940-X}.

\bibitem[{\citenamefont{Cokelaer}(2007)}]{Cokelaer:2007kx}
\bibinfo{author}{\bibfnamefont{T.}~\bibnamefont{Cokelaer}},
  \bibinfo{journal}{Phys. Rev. D} \textbf{\bibinfo{volume}{76}},
  \bibinfo{pages}{102004} (\bibinfo{year}{2007}), \eprint{0706.4437}.

\bibitem[{\citenamefont{Schutz}(1991)}]{schutz-91}
\bibinfo{author}{\bibfnamefont{B.}~\bibnamefont{Schutz}}, in
  \emph{\bibinfo{booktitle}{The Detection of Gravitational Waves}}, edited by
  \bibinfo{editor}{\bibfnamefont{D.}~\bibnamefont{Blair}}
  (\bibinfo{publisher}{Cambridge University Press},
  \bibinfo{address}{Cambridge, U.K.; New York, U.S.A.}, \bibinfo{year}{1991}),
  pp. \bibinfo{pages}{406--452}.

\bibitem[{\citenamefont{Press et~al.}(2007)\citenamefont{Press, Teukolsky,
  Vetterling, and Flannery}}]{NRecipes}
\bibinfo{author}{\bibfnamefont{W.~H.} \bibnamefont{Press}},
  \bibinfo{author}{\bibfnamefont{S.~A.} \bibnamefont{Teukolsky}},
  \bibinfo{author}{\bibfnamefont{W.~T.} \bibnamefont{Vetterling}},
  \bibnamefont{and} \bibinfo{author}{\bibfnamefont{B.~P.}
  \bibnamefont{Flannery}}, \emph{\bibinfo{title}{Numerical Recipes 3rd Edition:
  The Art of Scientific Computing}} (\bibinfo{publisher}{Cambridge University
  Press}, \bibinfo{year}{2007}).

\bibitem[{\citenamefont{Ajith}(2008)}]{Ajith:2007xh}
\bibinfo{author}{\bibfnamefont{P.}~\bibnamefont{Ajith}},
  \bibinfo{journal}{Class. Quant. Grav.} \textbf{\bibinfo{volume}{25}},
  \bibinfo{pages}{114033} (\bibinfo{year}{2008}), \eprint{arXiv:0712.0343
  [gr-gc]}.

\bibitem[{\citenamefont{Ajith et~al.}(2010)\citenamefont{Ajith, Bose, and
  Ghosh}}]{ABGNetSystErr}
\bibinfo{author}{\bibfnamefont{P.}~\bibnamefont{Ajith}},
  \bibinfo{author}{\bibfnamefont{S.}~\bibnamefont{Bose}}, \bibnamefont{and}
  \bibinfo{author}{\bibfnamefont{S.}~\bibnamefont{Ghosh}}
  (\bibinfo{year}{2010}), \bibinfo{note}{in preparation}.

\bibitem[{\citenamefont{Bose et~al.}(2008)\citenamefont{Bose, Gopakumar, and
  Tessmer}}]{Bose:2008ix}
\bibinfo{author}{\bibfnamefont{S.}~\bibnamefont{Bose}},
  \bibinfo{author}{\bibfnamefont{A.}~\bibnamefont{Gopakumar}},
  \bibnamefont{and} \bibinfo{author}{\bibfnamefont{M.}~\bibnamefont{Tessmer}}
  (\bibinfo{year}{2008}), \eprint{0807.2400}.

\end{thebibliography}

\end{document}